\documentclass[english]{article}
\usepackage[T1]{fontenc}
\usepackage[latin9]{inputenc}
\usepackage{graphicx}
\usepackage{esint}

\makeatletter
\newcommand{\lyxaddress}[1]{
\par {\raggedright #1
\vspace{1.4em}
\noindent\par}
}
\newenvironment{lyxlist}[1]
{\begin{list}{}
{\settowidth{\labelwidth}{#1}
 \setlength{\leftmargin}{\labelwidth}
 \addtolength{\leftmargin}{\labelsep}
 }}
{\end{list}}

\makeatother

\usepackage{babel}
\begin{document}

\title{Molecular emulsions: from charge order to domain order}

\author{Aurélien Perera}
\maketitle

\lyxaddress{Laboratoire de Physique Théorique de la Matière Condensée (UMR CNRS
7600), Université Pierre et Marie Curie, 4 Place Jussieu, F75252,
Paris cedex 05, France.}
\begin{abstract}
Aqueous mixtures of small molecules, such as lower n-alkanols for
example, are known to be micro-segregated, with domains in the nano-meter
range. One consequence of micro-segregated domains would be the existence
of long range domain-domain oscillatory correlations in the various
atom- atom pair correlation functions, and subsequent pre-peaks in
the corresponding atom atom structure factors, in the q-vector range
corresponding to nano-sized domains. However, no such pre-peak have
ever been observed in the large corpus of radiation scattering data
published so far. Here, through large scale simulations of aqueous-1propanol
mixtures, I report that the domain pre-peak contributions in the atom-atom
structure factors exactly cancel each other in the total scattering
intensity, thus suppressing the pre-peak in agreement with the experimental
findings. This cancellation is explained by drawing an analogy between
the charge order found in ionic fluids and the segregated domain order.
This finding opens new interpretation of the well known scattering
pre-peak observed in micro-emulsions. In particular, it implies that
scattering experiment cannot detect homogeneous domain segregation,
hence cannot lead to a proper microscopic description of atom-atom
correlations in domain ordered mixtures.
\end{abstract}

\section{Introduction}

Aqueous mixtures of small quantities of tbutanol or n-alkyl polyglycol
ether ($C_{n}E_{m}$) show aggregation of these solute molecules in
both cases\cite{101berne,102soperTBA0,103ani1,110NIS,111jose,112taillandier}.
However, there is a considerable physico-chemical difference between
these two types of aggregation. While tbutanol molecules form small
aggregates about 1nm wide\cite{103ani1}, $C_{n}E_{m}$ molecules
self-assemble into shapes called micelles about 5-10nm wide\cite{111jose}.
The latter type of mixtures are called emulsions, while the former
is a solution. In both cases, it is the dual amphiphilic/hydrophobic
nature of the two types of molecules that produces the aggregate formation\cite{200Tanford}.
The microscopic structural differences between these two mixtures
can be probed by radiation scattering\cite{201bernePecora,201book},
such as light, Xray or small angle neutron scattering. The intensity
$I(k)$ scattered off the micro-emulsion will show a Teubner-Strey
behaviour\cite{202TS} with an important pre-peak for wave vector
$k_{P}$ in the range $k_{P}\approx0.1-0.2$ \AA$^{-1}$, while for
aqueous tbutanol mixture $I(k)$ will show a typical Ornstein-Zernike
like behaviour, with no such pre-peak. The existence of the pre-peak
is only weakly dependent of the nature of the scattering radiation
type (light, neutrons, Xray), and seems to depend more on the type
of aggregates\cite{201bernePecora,201book}. The scattering data in
both systems seems to suggest that the existence of a pre-peak could
be related to the size of the aggregates. 

Since a decade, however, a new aspect of the solution type mixtures
has become apparent: the atom-atom pair correlation functions, \emph{as
obtained in computer simulations}, exhibit long range oscillations,
which come from the existence of correlations between aggregated domains\cite{203myIUPAC}.
Consequently, the corresponding atom-atom structure factors exhibit
a domain pre-peak. This finding poses a problem: why this pre-peak
does not contribute to the radiation scattered intensity $I(k)$? 

One of the possible answers is that computer simulations could produce
artifacts in the long range correlations, due to approximate molecular
model, or statistical problems. However, this would be in variance
with the fact that, for many neat liquids, computer simulations are
able to predict scattering pre-peak in excellent agreement with experiments,
for example neat alcohols\cite{203xMatija} and neat room temperature
ionic liquids\cite{203yMargulis,203zAlessandro}. The origin of these
pre-peaks has been related\cite{203yMargulis,203zAlessandro,204agilio,204aVoth}
to the segregation of the charged (hydroxyl or ionic) and neutral
(methyl or methylene) molecular groups. Furthermore, in such systems,
the contribution of pre-peaks of the atom-atom structure factors to
the total scattering pre-peak has been demonstrated. There is however
an important difference between these neat systems and the mixtures
mentioned above. As I have recently shown\cite{204MyChOrd}, the atom-atoms
correlations of these neat systems do not show any long range domain
oscillations corresponding to the segregation of the charged and neutral
groups. Hence, the pre-peaks in such systems do not arise from segregated
domain correlations, but correspond to the alternate disposition of
the plus and minus charges \emph{within} the charged group domain.
This observation shows the principal difference between those neat
systems which show scattering pre-peak and the case of mixtures presented
here. Indeed, in the present case, both segregated species have atomic
charges, such that, contrary to neat alcohols and ionic liquids, one
cannot speak of charged group versus uncharged group segregation.
What is even more problematic in the case of aqueous mixtures is precisely
the absence of pre-peak in scattering experiments, versus their existence
in atom-atom structure factors obtained from computer simulations.

The answer we provide here clarifies the origin of this discrepancy.
We show that these long range domain oscillations in the atom-atom
correlations are genuine physical features, but their contributions
to the radiation scattering intensity vanish, because the various
atom-atom contributions cancel each other. This cancellation is similar
to that found in simple ionic liquids, where the charge order imposes
out-of-phase long range oscillations\cite{204MyChOrd,204myIL}, hence
drawing an analogy between charge order and domain order. In addition,
the same analogy equally explains the existence of a pre-peak in the
scattering in micellar systems, through a mechanism similar that produces
a scattering pre-peak in room-temperature ionic liquids, as compared
to its absence in ordinary ionic liquids: the perturbation of the
charge order by the uncharged methyl groups\cite{204MyChOrd,204myIL}.
The equivalent mechanism in the case of domain order would be the
perturbation of this order by the large interface between the two
types of components. This new explanation gives a better microscopic
foundation to the previous argument based on size antagonism between
types of aggregates.

In the presentation below, we first recall the important details of
the charge ordering process in different types of ionic liquids. Then,
we present the case of aqueous-1pronanol as a prototype to show the
nature of the domain order and the consequences on both the various
atom-atom pair correlation function and the total scattering function.
In the final part, we examine the consequences of the domain and charge
order analogy for the understanding of the liquid-like order in complex
liquids.

\section{Charge ordering}

Charge ordering is a crucial concept for this paper. Since it has
been covered in a previous publications\cite{204MyChOrd,204myIL},
we will be brief about it in the presentation below. Charge ordering
describes the very special form of order in simple ionic liquids,
such as high temperature molten salts, for example. This special form
of order is apparent from the correlation functions between like and
unlike charge atoms, namely $g_{++}(r)$, $g_{--}(r)$ and $g_{+-}(r)$
, which are function of the atom-atom separation distance $r$. Fig.1
shows a typical example taken from a model simulation of a ionic liquid,
made of soft spheres of same diameter, which bear the charges of valence
$z_{+}=+1$ and $z_{-}=-1$. In Fig.1a, it is seen that, past the
details of the first neighbour correlations, the remainder of the
correlations are exactly out of phase. This property translates into
the following equality, which holds for large distances only, $r>r_{C}\approx3.6$\AA:

\begin{equation}
h_{++}=h_{--}=-h_{+-}\label{CO}
\end{equation}
where $h_{ij}=g_{ij}-1$. These equalities can be summarised in a
unique one as
\begin{equation}
\sum_{ij}h_{ij}=0\;\;\;\mbox{\ensuremath{r>r_{C}}}\label{CO2}
\end{equation}
Correlation functions for uncharged atoms never obey this property,
and are usually more of less in phase at large distances. Charge order
is therefore a remarkable form or order in a \emph{disordered} liquid.
The origin of this order is naturally coming from the fact that like
charges repel each other, while unlike charges attract each other,
and in a disordered liquid, these local constraint leads to this special
form of order. Charge order can equally be defined through the atom-atom
structure factors, which are related to the Fourier transform of the
correlation functions\cite{205Hansmac}:

\begin{equation}
S_{ij}(k)=\delta_{ij}+\rho\sqrt{x_{i}x_{j}}\int d{\bf r}\,h_{ij}(r)\,\exp(i{\bf k}.{\bf r})\label{Sk}
\end{equation}
where $x_{i}$ is the mole fraction of species $i$, and $\rho=N/V$
is the number density defined as the total number of atoms $N$ in
the volume $V$. The functions $S_{++}(k)$, $S_{--}(k)$ and $S_{+-}(k)$
are shown in the inset of Fig.1b. Charge order is visible through
the exact opposition of the peaks (shown by the red arrow), at the
k-vector $k\approx1.67$\AA$^{-1}$, which corresponds to the period
of the long range oscillation in the $g_{ij}(r)$. The fact that the
charge order peaks are exactly opposite in sign comes naturally from
the equalities in Eq.(\ref{CO}).

The key information from charger order, which will be very helpful
to understand domain order, \emph{is the exact cancellation of the
structure factors charge order peaks}. This is highlighted through
the Bhatia-Thornton transformation\cite{206bhatia}, which holds only
for binary mixtures. It consists in defining 2 new microscopic densities,
the total local density $\rho_{N}({\bf r})=\rho_{+}({\bf r})+\rho({\bf r})$
and the charge density $\rho_{Z}({\bf r})=[z_{+}\rho_{+}({\bf r})+z_{-}\rho_{-}({\bf r})]/2$,
and introducing corresponding new structure factors $S_{AB}(k)=<\tilde{\rho}_{A}(k)\tilde{\rho}_{B}(-k)>$
through ensemble averages of the correlations of their Fourier transforms.
In particular, one has for the density-density structure factor

\begin{equation}
S_{NN}=\frac{1}{2}\left[S_{++}+S_{--}+2S_{+-}\right]\label{BT}
\end{equation}
This structure factor $S_{NN}$ is equally represented (in orange)
both in the main panel (c) and in the inset of Fig.1. The capital
information from the main panel (c) is that the opposing peaks in
the $S_{ij}$ do not appear in $S_{NN}$, due to their exact cancellation
in the expression Eq.\ref{BT}. Conversely, these peaks appear in
the charge-charge structure factor shown in the inset of panel (c).
This cancellation crucially indicates that the density-density structure
factor, which is actually the observable, does not contain the information
about charge order. Hence, this structure factor looks like that of
a random mixture, totally conceiling the fact that there is an underlying
charge ordering beneath. In the next section, we will invoke a similar
analogy for domain ordering, which will explain why domain order pre-peaks
cancel in a similar way, which is the main message of this paper.

It is important to note that charge order is different from the global
electroneutrality, although both are obviously related through the
Coulomb interaction. Global electroneutrality is contained in the
small-k limit of the structure factors, through the well-known Stillinger-Lovett
sum rules\cite{206ST}. So they concern $k=0$ behaviour of the structure
factor. In contrast, charge order concerns the local distribution,
as witness by both the medium-to-long range oscillations and the $k\neq0$
wave vector where it manifests itself, and it may not necessarily
obey electroneutrality, which is a global ($k=0$) constraint.

The various features of the charge ordering process shown here, are
now used to demonstrate how domain-ordering follows a similar pattern
to charge ordering.

\section{Domain ordering in aqueous 1propanol mixtures}

We have studied by computer simulations the aqueous 1propanol mixtures,
and in particular various atom-atom correlation functions and corresponding
structure factors. This type of mixture corresponds to what we have
named molecular emulsions\cite{203myIUPAC}, which show strong micro-heterogeneity,
with water and solute segregated domains\cite{207ourTBA,208ourACE2}.
SPC/E water model\cite{300spce} and TraPPe 1propanol model\cite{301trappe}
were chosen. The focus is the 30\% 1propanol aqueous mixture, since
it is close to the maximum of the experimental Kirkwood-Buff integrals\cite{302Matteoli,303MyKBI},
where maximum segregation effects are expected. The structure of this
mixture has been previously studied by Xray and small angle neutron
scattering experiments\cite{303exp0,303exp1,303expt0,303expt2} as
well as computer simulations\cite{303simu1,303simu2}, and both approaches
revealed the clustering properties of these mixtures. It is important
to note that none of these works have reported the existence of scattering
or atom-atom domain correlation pre-peaks. The present simulations
have been conducted in the isobaric ensemble by using the gromacs
package\cite{304gromacs}. The temperature was maintained at 300K
through a Nosé-Hoover thermostat, and the pressure was let at 1atm
using the Parrinello-Rahman barostat, with time constant 1ps. Various
system sizes were investigated (see below). In each case, the system
was equilibrated for 5ns, and production runs for 10ns. In order to
properly sample long range oscillations due correlations between segregated
domains, we have studied a system of $N=128\:000$ molecules, which
corresponds to a box size of $L=195$\AA. This is an unusually large
number, but it is required, since lower system sizes do not allow
a proper sampling of these domain-domain correlations -as shown further
below. 

Fig.2a shows all the atom-atom correlation functions $g_{ab}(r)$
(where $a$ and $b$ stands for the various atoms) for the aqueous
mixture with 30\% 1propanol. These are the 3 functions $g_{O_{W}O_{W}},$$g_{O_{W}H_{W}}$
and $g_{H_{w}H_{w}}$ shown in blue for water (where the index $O_{W}$
and $H_{W}$ designate the water oxygen and hydrogen atoms, respectively),
the 15 functions $g_{OO}$ , $g_{OH}$ , $g_{OM_{1}}$ , $g_{OM_{2}}$
, $g_{OM_{3}}$ , $g_{OO}$ , $g_{OH}$ , $g_{OM_{1}}$ , $g_{OM_{2}}$
,$g_{OM_{3}}$, $g_{M_{1}M_{1}}$ , $g_{M_{1}M_{2}}$ , $g_{M_{1}M_{3}}$
,$g_{M_{2}M_{2}}$ , $g_{M_{2}M_{2}}$ , $g_{M_{2}M_{3}}$and $g_{M_{3}M_{3}}$
shown in green or 1propanol (where the index $O$, $H$, $M_{1}$,
$M_{2}$ and $M_{3}$ designate the 1propanol oxygen, hydrogen and
methylene/meythyl atoms, respectively), as well as the 15 cross correlation
functions (shown in magenta) between these atoms, for a total of 33
functions.

The short and long range parts of the functions are displayed in separate
vertical scales, in order to emphasize the different types of oscillations:
atom-atom correlations at short range and domain-domain correlations
at long range. In Fig.2a, the short range part (left panel) shows
all the distinct features due to specificities of the various atoms.
However, at long range(right panel), $r>r_{C}\approx25$\AA, all
the specificity of the atomic features disappear and merge into 3
distinct features which depend only in the species-species correlations,
corresponding respectively to the water-water, 1propanol-1propanol
and water-1propanol correlations. It is clearly seen that these 3
type of species-species correlations show out-of-phase correlations
between the like and cross correlations. The long range part of Fig.2a
bears a striking resemblance with charge-ordering displayed in Fig.1,
and we will consider here that these out of phase oscillations represent
a ``domain ordering'', which is due to the micro-segregation of
water and 1propanol(see snapshots in Fig.3).

Fig.2b shows the structure factors $S_{ab}(k)$ corresponding to the
correlation functions $g_{ab}(r)$ shown in Fig.2a. The inset shows
the domain pre-peak region enlarged. The pure 1propanol pre-peak is
indicated by a red arrow, and is seen to occur at the k-vector larger
than the domain pre-peak. Once again, we see that all the differences
in the atomic details shows up essentially for $k>k_{C}=2\pi/r_{C}\approx3.6$\AA$^{-1}$,
but for smaller k-values $k<k_{C}$ , only the species-species specificity
emerge into pre-peaks with opposing signs.. This finding proves that
the domain-domain correlations are independent of the atomic details
of the various molecular constituents. We have previously discussed
such pre-peaks in the context of many other types of aqueous mixtures,
and also non-aqueous mixtures\cite{203myIUPAC}. But it is the first
time that we relate such pre-peak to charge order, as seen in Fig.1,
which they are clearly reminiscent of. 

The micro-segregation of this system is illustrated in Fig.3, through
snapshots for the 30\% mixture of various sizes $N=20$00, $N=16000$
and $N=128000$ (each system is exactly the double of the size of
the previous one). In all these three cases, the local segregation
of domains is quite obvious, and these domains show an alternated
dispositions, which strikingly resembles that of the charge ordering
in the ionic liquid of the previous section, as shown in Fig.1a. There
are important differences though. While the charges in the ionic liquid
are localised within the atoms, the water and 1propanol domains do
have have such sharp localisation. This feature has important heuristic
implications that we will discuss later in Section 5. Although this
``domain order'' is much more loose than the strict charge order
of the ionic liquid, the long range correlations hold an appealing
analogy through the fact they appear to obey out-of-phase behaviour
reminiscent of that captured through Eqs.(\ref{CO},\ref{CO2}) for
the case of ionic liquids. This type of equality can be illustrated
further through the analysis of the atom-atom structure factors $S_{ab}(k)$,
shown in Fig.2b. 

The domain order displayed in Figs.2 has also a striking similarity
with the so-called bi-continuous and plumber phases found in micro-emulsions\cite{305bicont},
which exhibit segregation of molecular species at a larger scales,
but which occurs more often in presence of an water-oil-surfactant
context. The important difference that these have with the actual
system is the presence of better defined interface area, separating
water-rich and oil-rich domains, and which is saturated with the surfactant\cite{305bicont}.
We conjecture that it is this segregation of a component into a lower
dimensional area, which is responsible to the scattering Teubner Strey
pre-peak\cite{202TS} observed in such systems. We revisit this argument
below in Section 5.

The system size dependence is further illustrated in Fig.4a-b, where
we show the 3 oxygen atom correlation functions $g_{O_{W}O_{W}}$,
$g_{OO}$ and $g_{O_{W}O}$ (Fig.4a) , as well as the corresponding
structure factors (Fig.4b), calculated for the different system sizes
shown in Fig.3. In both Fig.4a and Fig.4b, the insets focus on the
respective domain-domain contributions. These figures show that the
system size does not matter so much for the short range features.
However, both insets show the dramatic differences coming from the
long range part, which is sensitive to correct description of domain-domain
correlations. For example, system size $N=2000$ leads to incorrect
and too large $k=0$ predictions of the structure factors. This well
known problem has been previously reported by us\cite{207ourTBA,208ourACE2}.
The $N=16000$ system seems appropriate since it gives results nearly
similar to the $N=128000$ system, although the r-range does not extend
beyond $48$AA.

\section{Absence of radiation scattering pre-peak in domain-ordered systems}

\subsection{Expression for the scattered intensity}

One of the problem of predicting domain pre-peaks in the atom-atom
structure factors is to explain why such pre-peaks are not experimentally
observed in radiation scattering experiments\cite{203myIUPAC}. We
provide the answer here. The radiation scattering intensity $I(k)$
is formally defined through the Debye formula\cite{305Debye}
\begin{equation}
I(k)=<\sum_{i,j}f_{i}(k)f_{j}(k)\exp\left(i{\bf k}.({\bf r}_{i}-{\bf r}_{j})\right)>\label{Idebye}
\end{equation}
where the sum runs over all pairs of scattering atoms i,j, which are
at respective spatial positions ${\bf r}_{i}$ and ${\bf r}_{j}$,
the functions $f_{i}(k)$ are the atomic form factor for atom $i$
and depend on the type of radiation which is scattered, and the symbol
<...> designates an average over all possible positions of these atoms,
which corresponds to a thermal average, or an ensemble average for
calculational purposes. In practice, it is convenient to rewrite this
expression in terms of the molecular species which contains the atoms\cite{306PW}.
For a binary mixture, we introduce symbols $i,j$ to designate the
molecular species index, and $a_{i},b_{j}$ to designate the atoms
of types a and b in respective molecules. Using the definition of
the atom-atom structure factor :

\begin{equation}
\rho\sqrt{x_{i}x_{j}}S_{a_{i}b_{j}}^{(M)}(k)=<\sum_{m_{a_{i}}m_{b_{j}}}\exp\left(i{\bf k}.({\bf r}_{m_{a_{i}}}-{\bf r}_{m_{b_{j}}})\right)>\label{S-k}
\end{equation}
where the sum runs over all atoms of type $a_{i},b_{j}$ , and $x_{i}=N_{i}/N$
is the mole fraction of molecular species $i$. In the equation above,
the atom-atom structure factor $S_{ab}^{(M)}(k)$ is the not the same
as that which appears in Eq.(\ref{Sk}), since it contains contributions
from the intra-molecular contributions as well, hence the superscript
(M) for molecular. Indeed, the sum in Eq.(\ref{S-k}) contains also
atom pairs in the same molecule. It can be shown, in case of atoms
rigidly bound inside a molecule, that this contribution in Eq.(\ref{S-k})
comes down to the Bessel function $j_{0}(kd_{ab})=\sin(kd_{ab})/kd_{ab}$,
where $d_{ab}=|{\bf r}_{a}-{\bf r}_{b}|$. This function is the same
as the W-matrix, with elements $w_{ab}(k)=j_{0}(kd_{ab})$, which
appears in the Site-Site Ornstein-Zernike theory\cite{205Hansmac},
and which contain the intra-molecular contribution to the pair correlation
function. The link with the structure factor defined through Eq.(\ref{Sk})
and the atom-atom pair correlation function $g_{a_{i}b_{j}}(r)$ is
then 

\begin{equation}
S_{a_{i}b_{j}}^{(M)}(k)=w_{a_{i}b_{j}}(k)+\rho\sqrt{x_{i}x_{j}}\int d{\bf r}\left[g_{a_{i}b_{j}}(r)-1\right]\,\exp(i{\bf k}.{\bf r})\label{SkM}
\end{equation}
 which represents a generalisation of Eq.(\ref{Sk}) to molecular
systems. By noting that the form factors $f_{i}(k)$ do not depend
on the thermal average < . >, we can rewrite the Debye formula into:

\begin{equation}
I(k)=\rho\sum_{ij}\sqrt{x_{i}x_{j}}\sum_{a_{i}b_{j}}f_{a_{i}}(k)f_{b_{j}}(k)S_{a_{i}b_{j}}^{(M)}(k)\label{PWM}
\end{equation}
which is convenient to recalculate the scattered intensity from the
atom-atom pair correlation functions and the corresponding atom-atom
structure factors. It is interesting to note that this expression
is similar to the Pings-Waser (PW) formula\cite{306PW} generically
used by many authors, but which does not contain the intra-molecular
part $w_{ab}(k)$. The present derivation shows both the origin of
this term and how to incorporate this contribution into the PW expression
through the correct expression Eq.(\ref{SkM}). The strict PW formula
is recovered by setting $W_{ab}(k)=\delta_{ab}$. The expression in
Eq.(\ref{PWM}) applies both for Xray and neutron scattering, when
appropriate form factors are used. In the case of neutron scattering,
$I(k)$ represents only the incoherent part of the scattering.

\subsection{Application to the aqueous-1propanol mixtures}

We now compute the Xray scattering intensity from various atom-atom
pair correlation functions and structure factors shown in Fig.2. The
form factors are taken from the scattering data\cite{307FormFactor}.
Fig.5 shows the total $I(k)$, as well as the 3 species-species contributions,
namely water-water (blue), propanol-propanol (green) and water-propanol
(magenta) contributions to $I(k)$. The dashed red line represent
the negative of the sum of water-water and propanol-propanol contributions,
which should match the magenta curve if exact cancellation should
occur, which is seen to be the case in the pre-peak region. From the
main panel of Fig.5, it is clearly seen that each of these contributions
in the pre-peak region are 2 orders of magnitude larger than the total
$I(k)$. However, the total contribution totally cancels the pre-peak,
as can be seen in the expanded view of $I(k)$ reported in the top
inset. This inset shows that only the 1propanol and water main peak
are dominant, at $k\approx0.65$\AA$^{-1}$ and $k\approx1.45$\AA$^{-1}$,
respectively. This cancellation is a striking result for several reasons.
First of all, it is consistent with the known absence of pre-peak
in the experimental Xray scattering data for this particular system
Ref.\cite{303expt2}. Secondly, in order for the cancellation of such
huge pre-peak contribution to happen, despite the fact that experimental
form factors are used in conjunction with calculated structure factors,
there must be a fine tuned adjustment of these cancellations. This
fact proves that domain order is a very robust physical phenomena.

Then, if one considers the relative good agreement between the calculated
$I(k)$ and the experimental one, as shown in the lower inset of Fig.5,
and considering the fact that the form factors are taken from the
experimental data, this agreement enforces the hypothesis that the
simulated atom-atom structure factors must be close to the experimental
data - if such data could be measured. Indirectly, it confirms that
the classical model representation must be good enough to provide
the observed cancellation. Finally, it is important to realise that
the absence of the pre-peak in the experimental data does not allow
to infer the existence of domain correlations in each of the partial
contributions, since these cannot be observed individually. This is
a dramatic finding, since it shows that the existence of the domain-domain
correlations can only be predicted from theory and apparently against
experimental observation.

The lower inset of Fig.5 shows a comparison between the calculated
$\Delta I(k)$ and the Xray data from Ref.\cite{303expt2} (shown
in blue). It is seen that the agreement is quite fair, including in
the pre-peak region. The agreement is better on the various peak positions
than in the data itself, which implies that the size of the molecules
are well described by the models but their distribution is slightly
dephased with respect to real one. The data reported in Ref.\cite{303expt2}
is $\Delta I(k)=I(k)/I_{Id}(k)-1$, where $I_{id}(k)$ corresponds
to the ideal part of the expression in Eq.(\ref{PWM}), in other words
when the structure factors $S_{ab}(k)$ in Eq.(\ref{SkM}) are replaced
by the first term in this equation, namely $w_{ab}(k).$ 

We have equally computed the neutron scattered intensity by using
the appropriate form factors for deuterated water and 1propanol, namely
$D_{2}O$ and $ODC_{3}H_{7}$, and the cancellation of the domain
pre-peaks occurs once again. The total I(k) looks very much like that
reported in the upper inset of Fig.5.

It is interesting to see how domain order affect the scattered intensity.
For this, we select in Eq.(\ref{PWM}) the k-vectors under the pre-peak
contributions, where only the species-species contributions are seen
and all atom-atom details are washed out. For this range of k-vectors,
the various atom-atom structure factors of a given species pair are
strictly similar:

\begin{equation}
S_{a_{i}b_{j}}(k)=S_{ij}(k)\;\;\;\mbox{for }0<k<k_{D}\label{k-range}
\end{equation}
where $k_{C}$ is the maximum k-vector for which the domain order
pre-peak is numerically distinctively defined. For this k-vector range,
the Debye expression in Eq.(\ref{PWM}) becomes

\begin{equation}
I(k)=\rho\sum_{ij}\sqrt{x_{i}x_{j}}F_{i}(k)F_{j}(k)S_{ij}(k)\;\;\;\mbox{for }0<k<k_{C}\label{Ik-range}
\end{equation}
where the effective form factor functions $F_{i}(k)$, which depend
now on species, rather than atoms are defined as:

\begin{equation}
F_{i}(k)=\sum_{a_{i}}f_{a_{i}}(k)\label{FF}
\end{equation}
This expression is very similar to the form factor one would get in
case of approximating a super-atom made of all the atoms inside a
single molecule CITE. This type of transformation is used to describe
the methyl and methylene group as super atoms, where the form factor
of the united atom is approximated as $f_{M}=f_{C}+nf_{H}$, and $f_{C}$
and $f_{H}$ are the form factors of the carbon and hydrogen atoms,
respectively, with $n=2,3$ for the methylene and methyl pseudo atoms,
respectively. 

What Eq.(\ref{Ik-range}) tells us is that pre-peak k-vector region
is dominated by scattering of the molecular species as pseudo-atoms,
indifferently to the atomic details. In that, it is strictly similar
to the ionic liquid model, with only 2 mono-atomic species, with the
appropriate pseudo-atom form factor $F_{i}(k)$. It is interesting
to see that, for a binary component, the expression in Eq.(\ref{Ik-range})
is similar to the Bathia-Thornton structure factor $S_{NN}$ in Eq.(\ref{BT}),
with $I(k)=\rho\left[F_{1}^{2}S_{11}+F_{2}^{2}S_{22}+2F_{1}F_{2}S_{12}\right]$,
for $0<k<k_{D}$. 

\section{Discussion}

Perhaps the most important consequence of domain order is the inability
of predicting this type of order from experimental scattering methodology,
as confirmed by the various experimental findings of the aqueous-1propanol
mixtures\cite{303exp1,303expt2}. Indeed, if the experimental $I(k)$
should be used to obtain the individual atom-atom structure factors
and correlation functions, one would not be able to find the large
pre-peaks which correspond to the domain-domain correlations, since
this information is absent in $I(k)$. Since scattering data for $I(k)$
exist for several system of aqueous mixtures and other such system
exhibiting micro-heterogeneity, the tentative to obtain correlation
functions and compare them with simulation data are most probably
erroneous, at least in what concern the long range part. This deficiency
is more a fundamental problem than a technical one. Indeed, it means
that one has to infer the existence and the microscopic details of
molecular segregation solely from models and theory. Therefore, one
should seek a better theoretical description of domain segregation,
as well as its consequences. 

Micro-segregation is analogous to charge order, in that the segregated
domains are disposed in quasi alternate fashion -a checker board type
order, in order to maximize the segregation, without leading to full
phase separation. However, charge order concerns particle with fixed
shapes, and occurs at the level of the particles themselves. In contrast,
domain order concerns fuzzy molecular domains, with a certain degree
of cross mixing which depends on the nature of the interactions. This
is the reason why domain order is only observed in the long range
part of the pair correlation functions. One can imagine domain order
as being a smooth distortion of charged particles into fuzzy domain,
hence going from a particle representation into a field representation.
Strictly speaking, domain order is the field representation of charge
order.

There is an important difference concerning the valence of the particle
charges, which dictates charge order, and the domain valence, which
dictates domain ordering. Indeed, like-charges repel each other while
unlike-charges attract each other, leading to a large peak in the
cross correlation function. Conversely, in domain ordering, it is
the like correlations which exhibit the high peak at atomic contact,
while cross correlations are depressed at contact. In terms of interactions,
if a strict mapping with the Coulomb interactions should be made,
this behaviour would correspond to \emph{pure imaginary charges}.
If a ionic liquid with imaginary charges would be used, it will lead
to immediate demixing of each valence. Therefore, one requires a supplementary
mechanism to maintain the particles into charge order. This mechanism
is provided here through the hydrophobic-hydrophilic interaction balances,
which maintain the domain order. Moreover, domain ordering is seen
as a part of the atom-atom correlations, namely the long range part.
In order to describe this contribution, one could resort to a field
theory description, by assigning a phase to each of the fields corresponding
to atom-atom distribution functions.

It is interesting that the demonstration of domain order requires
extensive computer simulations, nearly at the edge of what can be
done in desktop PC-type workstations. Moreover, it seems reasonable
to think that most of soft-matter system experience such domain order
at various degrees of extent. In view of the computer power required
in the present case of aqueous-1propanol mixture, a system which cannot
be considered as particularly challenging, one wonders at the type
of resources that could be required to account for domain order in
simulation of realistic systems as those found in soft-matter. Conversely,
one wonders how much importance domain order can have in such system,
and to what extent it can be neglected. These are subjects for subsequent
investigations, which are newly opened by the present investigation.

When going from simple ionic liquids, such as molten NaCl, to room
temperature ionic liquids, such as ethylammonium-nitrate, for example,
the particles change from simple charged atoms to complex molecules,
which contain uncharged methyl and methylene groups\cite{40PP-IL1,40PP-IL2}.
These groups perturb the global homogeneity of the charge ordering
found in simple ionic liquids, and the loss of global homogeneity
gives rise to a pre-peak both in the atom-atom structure factors and
the total scattered intensity\cite{204MyChOrd,204myIL,40PP-IL1,40PP-IL2}.
Similarly, when going from molecular emulsions to micro-emulsions,
we conjecture that the global domain order is perturbed by the change
in the nature of the aggregates. Direct micelles contain extended
water impregnated outer interface, sharply separated from the inner
oily core\cite{110NIS}. Inverse micelles also have well separated
polar inner cores from the fuzzy and hairy outer core made of the
hydrophobic tails\cite{200Tanford}. This is a sharp change in the
nature of the type of aggregates found in molecular emulsions. We
conjecture here that is this change which produces a non canceling
pre-peak in the total scattered intensity. Demonstrating this conjecture
would require computer simulations beyond desktop capabilities. Yet,
many systems found in soft-matter or biomaterial context, contain
structural aggregates similar to micelles\cite{50scatt-1,50scatt-2,50scatt-3}.
The fact that scattering experiments in such systems are able to predict
scattering pre-peak\cite{50scatt-3} does not necessarily imply that
they can provide a better microscopic description. Indeed, there are
probably hidden cancellation mechanisms beneath the apparent pre-peak
of micro-emulsions, which would require further investigations similar
to that conducted herein. The microscopic relation between micelle
structure and scattering data remains to be re-investigated in the
light of the present finding. We conjecture that the present theoretical
descriptions of micelle formation, which never take into account domain
ordering, are analogous to the Debye-Huckel description of ionic liquids,
which does not account for charge order, but describes correctly the
screening of charges. A new theory of micro-emulsions, which accounts
for domain order, is yet to be developed. 

Another point concerns the actual physical impact of domain order
in terms of radiation scattering for the molecules themselves. In
other words, how does the fact that radiation scattering does not
witness a particular spatial frequency, despite the existence of domain
segregation, influences the actual physico-chemical properties of
the molecules themselves, specially if some of them are radiation
sensitive? The answer to this question could have a interesting consequences
in the bio-context, for example. The same question can be asked at
the more general point of view of any form of micro-segregated matter.

\section{Conclusion}

The most important conclusion of this work is the fact that micro-segregation
in molecular emulsions induces domain order, which is a form of macroscopic
homogeneity, rescaled at the level of the segregated domains as pseudo-molecular
grains. It is this rescaled homogeneity which is the principal reason
why scattering experiments cannot detect the underlying domain segregation.
This apparent homogeneity is similar to that found in atomic ionic
liquid, where the particles do not experience the random disordered
distribution seen in simple binary mixtures, but the charge order
dictated by the Coulomb interactions. It is this form of order-within-disorder,
which produces the apparent homogeneity of these systems. Micro-segregated
mixtures have the same type of order, hence they look more homogeneous
that they actually are, at least from the point of view of scattered
radiations. This apparent homogeneity can be destroyed by super-structures
such as micelles, which appear when going from molecular emulsions
to micro-emulsions. This is not a simple change in the size of aggregates,
as previously thought, but a topological change in the conformal structure
of the aggregate, which in particular induces a change in the homogeneity
of the system. Finally, we have emphasized that the data from scattering
experiments cannot lead to a complete atom-atom microscopic structure
of complex micro-heterogeneous liquid mixtures, in severe contrast
with the computer simulation calculations. 

\newpage

\section*{Figure captions}
\begin{lyxlist}{00.00.0000}
\item [{Fig.1}] (a) Snapshot of charge order for the ionic liquid characterized
by the parameters below the picture. (b) Pair correlation functions
and corresponding BT functions (inset; see text). (c) Structure factors
and corresponding BT functions (inset; see text). The red arrows indicates
the pre-peak position.
\item [{Fig.2}] (a) Atom-atom pair correlation functions of the aqueous-1propanol
30\% mixture. The functions are displayed into 2 different distance
scale (as indicated by the blue arrows) separating the short and long
range parts. the inset shows details of the correlations at atomic
contact. Blue is for water-water correlations, magenta for propanol-propanol
and green for cross correlations. The functions $g_{O_{w}O_{w}}(r)$,
$g_{O_{P}O_{P}}(r)$ and $g_{O_{W}O_{P}}(r)$, corresponding to the
correlations between the oxygen atoms of water and 1propanol, are
highlighted in thicker lines. (b) Corresponding atom-atom structure
factors with same line and color conventions. The red arrow indicates
the position of the neat propanol pre-peak. The inset represent a
zoom over the pre-peak part.
\item [{Fig.3}] Snapshots of the systems for various sizes: (a) N=2000,
(b) N=16000 and (c) N=128000, showing the segregation of the polar
OH groups and the oily methyl groups (in cyan). The water oxygen and
hydrogen atoms are shown in red and white, respectively, those of
1propanol in blue and grey, respectively, and the methyl/methylene
groups are shown in cyan.
\item [{Fig.4}] System size dependence of the atom-atom pair correlation
functions (a) and corresponding structure factors (b), illustrated
for the three oxygen-oxygen correlations, namely water-water (WW),
1propanol-1propanol (PP) and cross (WP). The data for system size
128000 is shown in full lines, 16000 in dotted lines and 2000 in dashed
lines. Corresponding WW correlations are colored in blue, cyan and
purple, respectively, for PP correlations in green, yellow and grass,
and cross WP correlations in magenta, red and brown, respectively.
In (b), the structure factor of pure water is shown as thin black
curve.
\item [{Fig.5}] Xray scattering intensity for the 30\% 1propanol in the
aqueous mixture, as computed from collecting the atom-atom structure
functions calculated in the simulations, through Eq.(\ref{PWM}).
The main panel shows the 3 partial species-species contributions to
I(k) (blue for water, green for propanol and magenta for cross) and
the calculated $I(k)$ in black (for the dotted red line, see text).
The calculated I(k) is reproduced in magnified scale in the upper
inset. The lower inset shows a comparison with experiments (shown
in blue from Ref.\cite{303expt2}) in blue of the quantities $\Delta I(k)=k\left[I(k)/I_{Id}(k)-1\right]$
(see text).
\end{lyxlist}
\newpage

.

\includegraphics[scale=0.25]{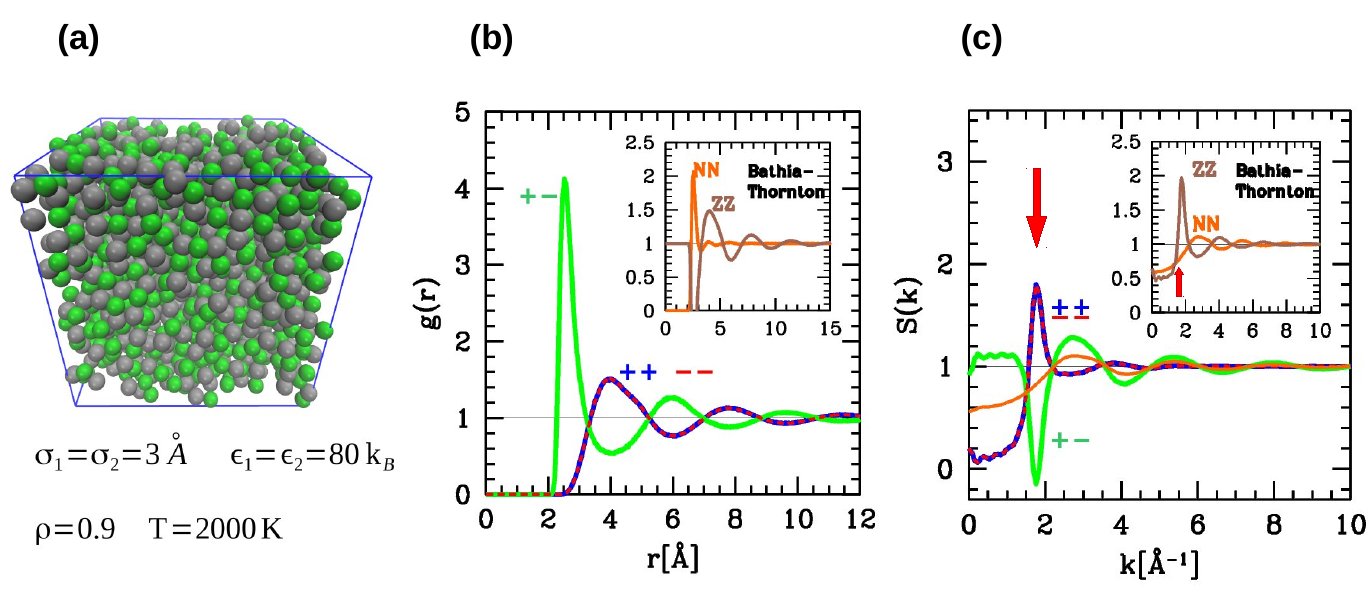}
.

Fig.1. (a) snapshot of charge order for the ionic liquid characterized
by the parameters below the picture. (b) Pair correlation functions
and corresponding BT functions (inset; see text). (c) Structure factors
and corresponding BT functions (inset; see text). The red arrows indicates
the pre-peak position.

.

\newpage

.

\includegraphics[scale=0.3]{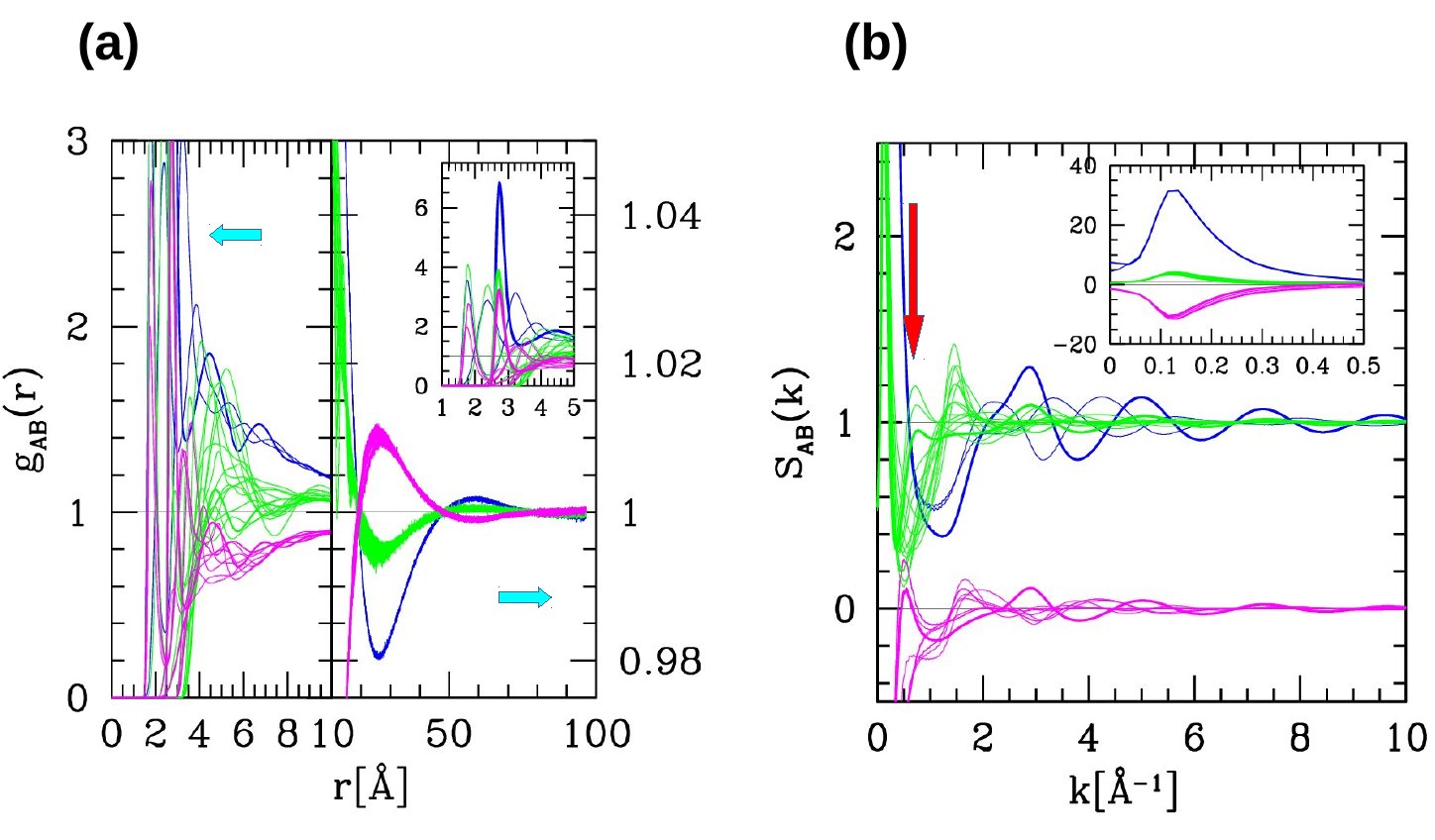}

.

Fig.2. (a) Atom-atom pair correlation functions of the aqueous-1propanol
30\% mixture. The functions are displayed into 2 different distance
scale (as indicated by the arrows) separating the short and long range
parts. the inset shows details of the correlations at atomic contact.
Blue is for water-water correlations, green for propanol-propanol
and magenta for cross correlations. The functions $g_{O_{w}O_{w}}(r)$,
$g_{O_{P}O_{P}}(r)$ and $g_{O_{W}O_{P}}(r)$, corresponding to the
correlations between the oxygen atoms of water and 1propanol, are
highlighted in thicker lines. (b) Corresponding atom-atom structure
factors with same line and color conventions. The inset represent
a zoom over the pre-peak part.

.

\newpage

.

\includegraphics[scale=0.35]{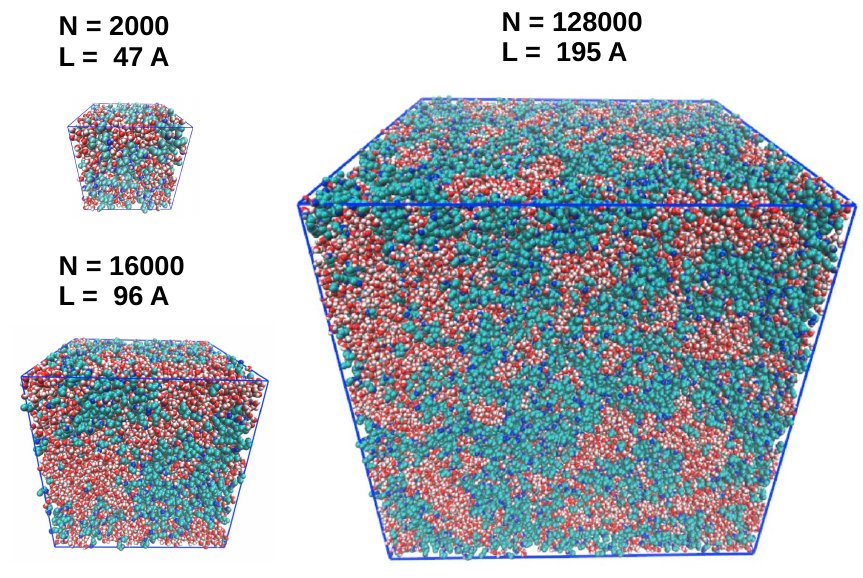}

.

Fig.3. Snapshots of the systems for various sizes: (a) N=2000, (b)
N=16000 and (c) N=128000, shown the segregation of the polar OH groups
and the oily methyl groups (in cyan). The water oxygen and hydrogen
atoms are shown in red and white, respectively, those of 1propanol
in blue and grey, respectively, and the methyl/methylene groups are
shown in cyan.

.

\newpage

.

\includegraphics[scale=0.3]{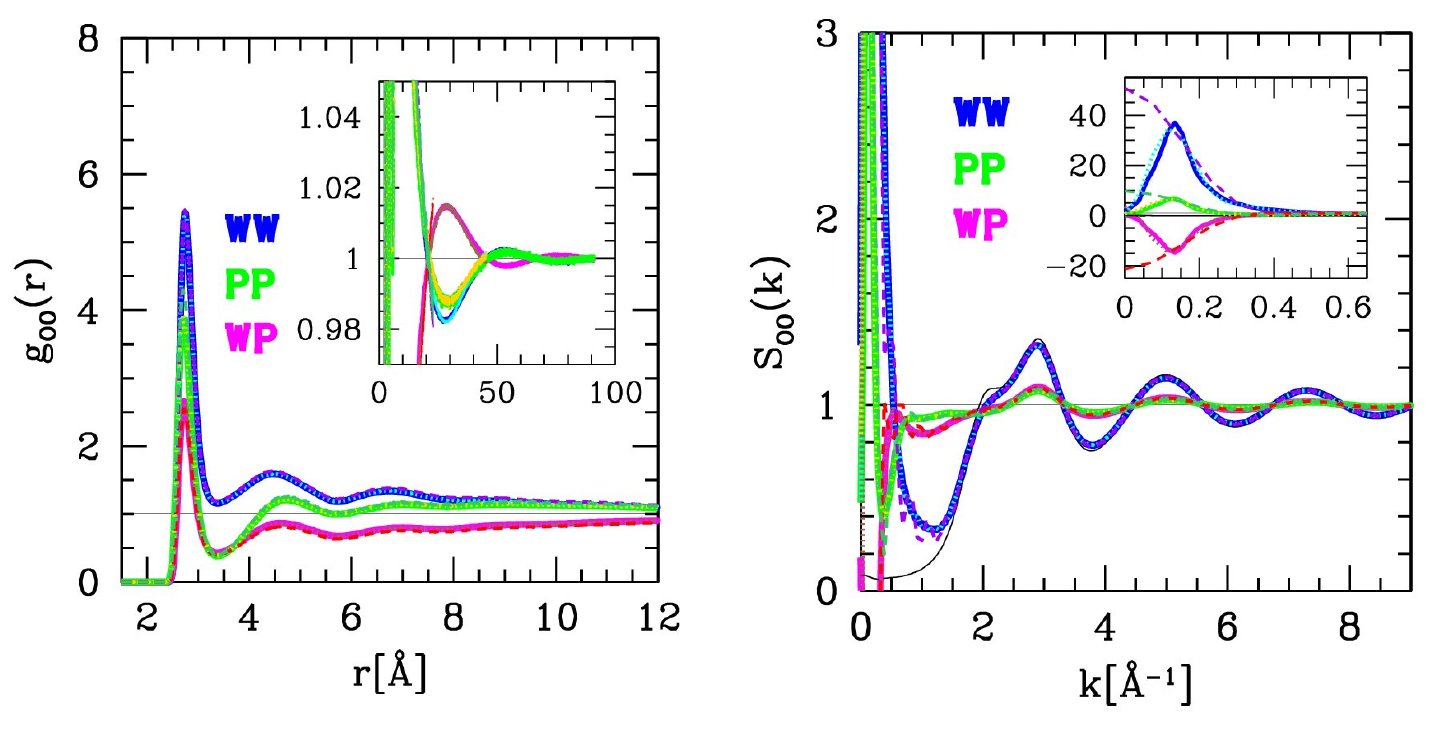}

.

Fig.4. System size dependence of the atom-atom pair correlation functions
(a) and corresponding structure factors (b), illustrated for the three
oxygen-oxygen correlations, namely water-water (WW), 1propanol-1propanol
(PP) and cross (WP). The data for system size 128000 is shown in full
lines, 16000 in dotted lines and 2000 in dashed lines. Corresponding
WW correlations are colored in blue, cyan and purple, respectively,
for PP correlations in green, yellow and grass, and cross WP correlations
in magenta, red and brown, respectively. In (b), the structure factor
of pure water is shown as thin black curve.

.

\newpage

.

\includegraphics[scale=0.4]{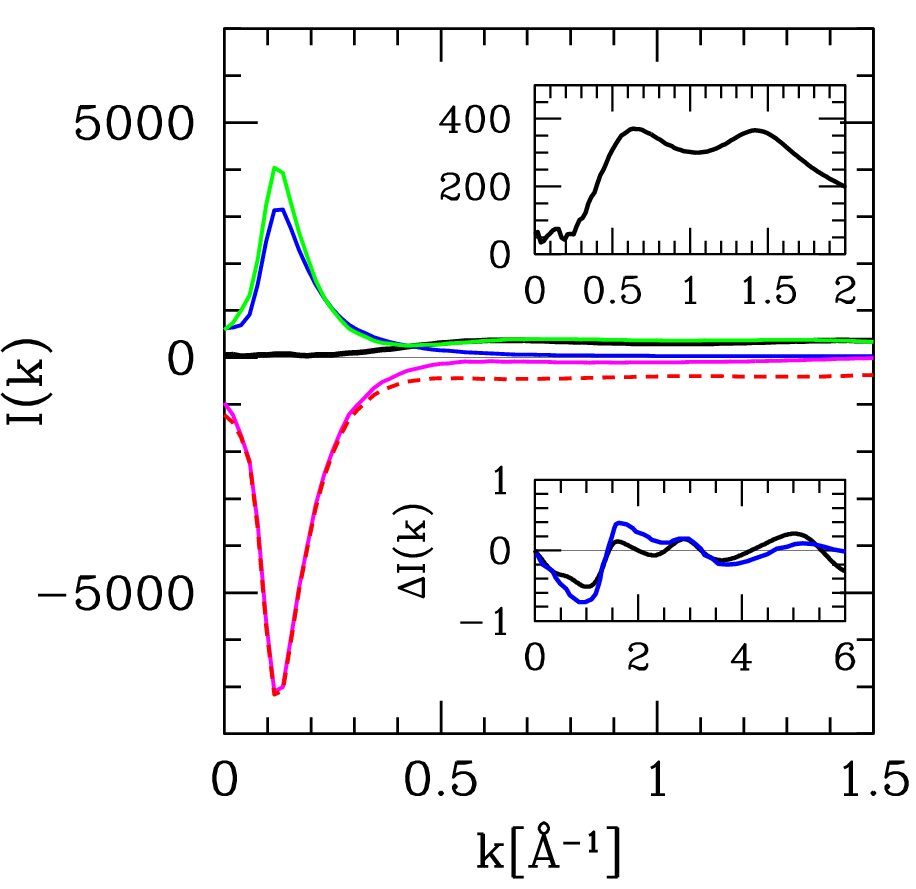}

.

Fig.5. Xray scattering intensity for the 30\% 1propanol in the aqueous
mixture, as computed from collecting the atom-atom structure functions
calculated in the simulations, through Eq.(\ref{PWM}). The main panel
shows the 3 partial species-species contributions to I(k) (blue for
water, green for propanol and magenta for cross) and the calculated
$I(k)$ in black (for the dotted red line, see text).The calculated
I(k) is reproduced in magnified scale in the upper inset. The lower
inset shows a comparison with experiments (shown in blue from Ref.\cite{303expt2})
of the quantities $\Delta I(k)=k\left[I(k)/I_{Id}(k)-1\right]$ (see
text).
\end{document}